\documentclass[11pt,reqno,a4,onecolumn,twoside,final,spanish]{article}

\usepackage{amsthm,amsmath,amssymb,latexsym,amsfonts,amscd} 
\usepackage{latexsym}
\usepackage{url}
\usepackage[T1]{fontenc}
\usepackage{ae}
\usepackage{aecompl}
\usepackage[small,sc,hang,bf]{caption} 
\usepackage{graphicx,color,ae,fancyvrb}
\usepackage{pst-all}
\usepackage{subfig}
\usepackage{float} 
\usepackage{verbatim}
\usepackage{enumerate}
\usepackage{array}
\usepackage{longtable}
\usepackage{multicol}
\usepackage{cancel}
\newcommand*\chancery{\fontfamily{pzc}\selectfont}

\newlength\mytemplen
\newsavebox\mytempbox

\makeatletter
\newcommand\mybluebox{%
    \@ifnextchar[
       {\@mybluebox}%
       {\@mybluebox[0pt]}}

\def\@mybluebox[#1]{%
    \@ifnextchar[
       {\@@mybluebox[#1]}%
       {\@@mybluebox[#1][0pt]}}

\def\@@mybluebox[#1][#2]#3{
    \sbox\mytempbox{#3}%
    \mytemplen\ht\mytempbox
    \advance\mytemplen #1\relax
    \ht\mytempbox\mytemplen
    \mytemplen\dp\mytempbox
    \advance\mytemplen #2\relax
    \dp\mytempbox\mytemplen
    \colorbox{myblue}{\hspace{1em}\usebox{\mytempbox}\hspace{1em}}}

\makeatother

\usepackage{natbib}

\usepackage{hyperref}

\usepackage{geometry}
\usepackage{calc}
	
\makeatletter
\renewcommand{\section}{\@startsection{section}{1}{0pt}{-3ex plus -1ex minus 0ex}{2ex plus 0ex}{\bf}}
\makeatother

\makeatletter
\renewcommand{\subsection}{\@startsection{subsection}{1}{0pt}{-2ex plus -1ex minus 0ex}{2ex plus 0ex}{\bf}}
\makeatother

\theoremstyle{definition}

\theoremstyle{remark}

\begin{document}

\renewcommand{\tablename}{Tabla}
\renewcommand{\figurename}{Figura}
\noindent

\begin{flushleft}
\textsl {\chancery  Memorias de la Primera Escuela de Astroestad\'istica: M\'etodos Bayesianos en Cosmolog\'ia}\\
\vspace{-0.1cm}{\chancery  9 al 13 Junio de 2014.  Bogot\'a D.C., Colombia }\\
\textsl {\scriptsize Editor: H\'ector J. Hort\'ua}\\
\href{https://www.dropbox.com/sh/nh0nbydi0lp81ha/AACJNr09cXSEFGPeFK4M3v9Pa}{\tiny {\blue Material suplementario}}
\end{flushleft}



\thispagestyle{plain}\def\@roman#1{\romannumeral #1}



\begin{center}\Large\bfseries Alternativa de aplicaci\'on para la radiaci\'on de fondo en el desarrollo de la base de datos del atlas de radiaci\'on atmosf\'erica \end{center}
\begin{center}\normalsize\bfseries  Alternative application for the radiation background in the development of the atlas database of atmospheric radiation \end{center}

\begin{center}
\small
\textsc{Ivan Arturo Morales De la Hoz \footnotemark[1]}
\footnotetext[1]{Fundaci\'on Universitaria Los Libertadores Bogot\'a, Colombia. E-mail: \url{iamoralesd@libertadores.edu.co}}

\end{center}

\noindent\\[1mm]
{\small
\centerline{\bfseries Resumen}\\
En la actualidad la radiaci\'on hace parte de las  variables a considerar dentro del marco de los pron\'osticos ambientales y es significativa en el incremento del calentamiento global, sumado al efecto invernadero. La radiaci\'on que hoy en d\'ia  contemplan los organismos meteorol\'ogicos dependen de la Referencia Radiom\'etrica Mundial (WRR), el Grupo Mundial de Normalizaci\'on (WSG), direccionada por el congreso de la Organizaci\'on Meteorol\'ogica Mundial (O.M.M.). La presente investigaci\'on se basa en la radiaci\'on c\'osmica de fondo  de microondas, como una variable a estimar para la toma de informaci\'on de la radiaci\'on incidente en la atmosfera terrestre,  con un aporte significativo en la construcci\'on del atlas de radiaci\'on por parte de la (UPME) e (IDEAM). Debido a que por el momento se contemplan variables de radiaci\'on ultravioleta e infrarroja, columna de ozono, radiaci\'on directa y radiaci\'on difusa,  las dos \'ultimas  resuelven la obtenci\'on de la radiaci\'on global, y se  eval\'uan por parte de los organismos meteorol\'ogicos del pa\'is. El estudio de la radiaci\'on de fondo como proyecto de investigaci\'on  arrojara  datos que robustecer\'an los sistemas de informaci\'on meteorol\'ogicos en Colombia, soportados a trav\'es de la metodolog\'ia Bayesiana. Demostrando que la radiaci\'on c\'osmica de fondo puede llegar a contribuir notoriamente en el aumento en la variabilidad de la temperatura troposfera, estratosfera y mesosfera mundial por  consecuencia del efecto invernadero. \\
{\footnotesize
\textbf{Palabras clave:}
Atm\'osfera terrestre, efecto invernadero, radiaci\'on c\'osmica de fondo. \\
\noindent\\[1mm]
{\small
\centerline{\bfseries Abstract}\\

Nowadays radiation is one of the variables to be considered in the environmental
forecasting and it is meaningful in the increase of global warming, together green-
house effect.  The radiation   considered by the meteorological organizations depends on the World Radiometric Reference (WRR), the World Standard Group (WSG), addressed by the World Meteorological Organization (WMO).  This work is based on the cosmic microwave background, as a variable to be estimated in order to get information about the incident radiation in the Earth`s atmosphere, as a valuable and meaningful contribution in the building of the radiation atlas by the (UPME) and (IDEAM). Due to the fact that  the variables considered are ultraviolet and infrared radiation, ozone column, direct radiation and diffuse radiation, the last two get the global radiation, and are the only ones to be evaluated by the national meteorological organizations in the country. The study of the cosmic background radiation as a research project will provide data which will strengthen the meteorological information systems in Colombia, supported by the Bayesian methodology; supporting that  cosmic background radiation can contribute significantly in the increase of the variability of the troposphere, stratosphere and mesosphere world temperature as a consequence of global warming. \\

{\footnotesize
\textbf{Keywords:}
Earth's atmosphere, greenhouse effect, cosmic background radiation.\\
}

\newpage
\section{Introducci\'on}
Los rayos c\'osmicos fueron descubiertos inesperadamente en 1912. Ahora se sabe que la mayor\'ia de los rayos c\'osmicos son n\'ucleos at\'omicos. La mayor\'ia son n\'ucleos de hidr\'ogeno; algunos son n\'ucleos de helio y elementos m\'as pesados que el resto. Los cambios de abundancia relativa con la energ\'ia de los rayos c\'osmicos son los m\'as altos de energ\'ia y tienden a ser n\'ucleos m\'as pesados. Aunque muchos de los rayos c\'osmicos de baja energ\'ia proviene de nuestro sol, el origen de los rayos c\'osmicos m\'as energ\'eticos sigue siendo desconocido y un tema de mucha investigaci\'on.  Los rayos c\'osmicos pueden incluso ser importantes para el clima de la Tierra, como el rayo com\'un que puede ser activado al pasar los rayos c\'osmicos, \cite{ivan1}.\\
\begin{figure}[h!]
\centering
\includegraphics[width=80mm]{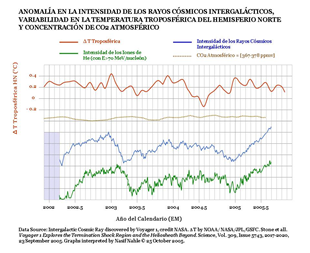}
\caption{Gr\'afico Cr\'edito: E. C. Stone, et al. Voyager 1 explora el choque de terminaci\'on y la Regi\'on Heliosheat all\'a. Ciencia; Vol. 309, pp 2017-2020. 23 de septiembre 2005. NOAA. }
\label{ivan1a}
\end{figure}
En la figura \ref{ivan1a}, se muestra  la superposici\'on sobre las oscilaciones en la temperatura de la troposfera para compararla con la gr\'afica sobre la Intensidad de los rayos c\'osmicos (IRCI) que colisionan con el viento solar en los confines del Sistema Solar. Parece demostrar que existe una correlaci\'on directa entre las variaciones de la temperatura troposf\'erica y la IRCI. Adem\'as, podr\'ia ser que la anomal\'ia de la IRCI provoque las anomal\'ias observadas en la actividad radiante de nuestro Sol, \cite{ivan2}.
\section{La anomal\'ia clim\'atica ant\'artica y los rayos c\'osmicos gal\'acticos} 
Se ha propuesto que los rayos c\'osmicos gal\'acticos pueden influir en el clima terrestre afectando la formaci\'on de nubes. Si los cambios en la nubosidad desempe\~nan un papel en el cambio clim\'atico, su efecto cambia de signo en la Ant\'artida. Datos del sat\'elite del Earth Radiation Budget Experiment (ERBE) se utilizan aqu\'i para calcular los cambios en la temperatura de la superficie en todas las latitudes, debido a peque\~nos cambios porcentuales en la nubosidad. Los resultados coinciden con los contrastes observados en los cambios de temperatura, a nivel global y en la Ant\'artida. Evidentemente las nubes no s\'olo responden pasivamente a los cambios clim\'aticos, sino que toman parte activa en el forzamiento, de conformidad con los cambios en el campo magn\'etico solar que var\'ian el flujo de rayos c\'osmicos, \cite{ivan3}.

\section{Cosmoclimatolog\'ia}

 Las nubes tienen un elevado albedo y ejercen su efecto de enfriamiento dispersando en el cosmos gran parte de la luz solar que de lo contrario podr\'ia calentar la superficie. Pero las nieves en las capas de hielo del Ant\'artico son asombrosamente blancas, con un mayor albedo que las nubes. All\'i, una cubierta de nubes extra calienta la superficie y menos nubosidad la enfr\'ia. Las mediciones por sat\'elite muestran el efecto del calentamiento de nubes sobre la Ant\'artida, y los meteor\'ologos en  regiones meridionales  lo confirman por observaci\'on. Groenlandia tambi\'en tiene una capa de hielo, pero es m\'as peque\~na y no tan blanca. Y mientras que las condiciones en Groenlandia est\'an acopladas al clima general del hemisferio norte, la Ant\'artida est\'a en gran medida aislada por torbellinos en el oc\'eano y en el aire.
La hip\'otesis de los rayos c\'osmicos y forzamiento de las nubes, predice que los cambios de temperatura en la Ant\'artida deben ser opuestos en signo a los cambios de temperatura en el resto del mundo. Esto es exactamente lo que se observa, en un fen\'omeno conocido que algunos geof\'isicos han llamado el sube y baja polar, para el que ``la anomal\'ia clim\'atica Ant\'artica'' parece un mejor nombre, \cite{ivan4}.

\section{Rayos c\'osmicos y el clima de la Tierra}
En la figura \ref{ivan2a}, hay una relaci\'on entre el enlace de los rayos c\'osmicos y la actividad solar para el clima terrestre. La actividad solar cambiante es responsable de la fuerza del viento solar que var\'ia. Un viento m\'as fuerte reducir\'a el flujo de rayos c\'osmicos que alcanzan la tierra, ya que una mayor cantidad de energ\'ia se pierde a medida que se propagan hasta el viento solar. Los mismos rayos c\'osmicos provienen desde afuera del sistema solar, lo m\'as probable es que sean acelerados por remanentes de supernova. Dado que los rayos c\'osmicos dominan la ionizaci\'on troposf\'erica, un aumento de la actividad solar se traducir\'a en una reducci\'on de la ionizaci\'on, como tambi\'en para una nube reducida de baja altitud. Dado que las nubes de baja altitud tienen un efecto neto de enfriamiento (su ``blancura'' es m\'as importante que su efecto ``manta''), el aumento de la actividad solar implica un clima m\'as c\'alido. Las variaciones del flujo de rayos c\'osmicos tienen un efecto similar, que no est\'an relacionadas con las variaciones de la actividad solar.
\begin{figure}[h!]
\centering
\includegraphics[width=80mm]{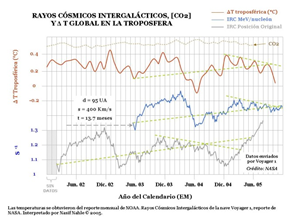}
\caption{Temperatura en la troposfera y rayos c\'osmicos intergal\'acticos.  Gr\'afico Cr\'edito: E. C. Stone, et al. Voyager 1 explora.}
\label{ivan2a}
\end{figure}
 En la figura \ref{ivan2a},  se ve claramente c\'omo la Intensidad de la Radiaci\'on C\'osmica intergal\'actica influye en la temperatura troposf\'erica terrestre. Cuando el viento solar choca con la radiaci\'on c\'osmica Intergal\'actica, los nucleones y el plasma de electrones del viento solar se calientan y disminuyen su velocidad de desplazamiento hacia afuera del Sistema Solar. En la terminaci\'on de choque, los electrones y nucleones de la radiaci\'on c\'osmica Interestelar penetran contracorriente por las ondas del viento solar y son desviados por la turbulencia magn\'etica que produce el movimiento del Sistema Solar desplaz\'andose hacia la terminaci\'on de choque. Los nucleones intergal\'acticos con baja densidad de energ\'ia no penetran el Sistema Solar sino que son desviados por los turbulencias magn\'eticas (Arco de Choque) que se forman por el impacto entre el Viento Solar y la RCI; sin embargo, las part\'iculas lentas con alta densidad de energ\'ia (part\'iculas calientes) remontan el viento Solar contra corriente, ellas se enfr\'ian de nuevo, y entonces  aceleraran hasta alcanzar velocidades supers\'onicas que alcanzan los 400 km/s viajando hacia el sol, es decir, en direcci\'on opuesta hacia la cual el Viento Solar fluye. La RCI y las part\'iculas aceleradas golpean contra el Campo Magn\'etico Terrestre (CMT). La colisi\'on de esas part\'iculas del arco de choque que colisionan en el CMT promueve la formaci\'on de nubes cuando penetran en la troposfera de la Tierra. Las part\'iculas de la RCI entrantes que inciden sobre la superficie de la Tierra incrementan la temperatura del suelo y de los oc\'eanos. El calor de la superficie se transfiere a la troposfera baja y \'esta se calienta. La intensidad de las part\'iculas intergal\'acticas y de la radiaci\'on c\'osmica que afectan a la Tierra depende de la intensidad del Viento Solar. Si la intensidad del Viento Solar es alta, entonces la RCI entrante desde el arco de choque del Sistema Solar ser\'ia m\'as alta tambi\'en. Si el Viento Solar disminuye su velocidad, la RCI que remont\'o el viento Solar contra corriente no disminuye su velocidad; sin embargo, las part\'iculas de la RCI no se desv\'ian, aunque ingresan a la Tierra, en donde transfieren su energ\'ia a las mol\'eculas del suelo y los oc\'eanos, calent\'andolos de forma extraordinaria. Si la actividad solar es intensa, entonces el flujo del plasma c\'osmico ser\'a mayor, \cite{ivan2}. La terminaci\'on de choque, o margen de Choque, es la regi\'on del Sistema Solar en donde las emisiones solares colisionan contra los rayos c\'osmicos Intergal\'acticos. Junto con la radiaci\'on del Sol, esta es la verdadera causa del Calentamiento Global y ha ocurrido c\'iclicamente durante toda la historia de nuestro planeta; algunas veces muy intensamente y en otras mucho menos. Ser\'ia muy poca la diferencia (un 3\% cuando mucho) si tuvi\'esemos la mitad de los gases de Invernadero en nuestra atmosfera. De acuerdo con las gr\'aficas incluidas arriba, en este momento estamos pasando por un per\'iodo de excesivamente alta actividad, tanto solar como Intergal\'actica. Los pron\'osticos no son muy buenos, pues estamos recibiendo toda la energ\'ia y los nucleones que se aceleraron, de tal forma que la temperatura troposf\'erica terrestre continuar\'a ascendiendo. No sabemos hasta cu\'ando ni cu\'anto; sin embargo, este ciclo terminar\'a y probablemente seguir\'a otro ciclo de enfriamiento global, el cual podr\'ia convertir una vez m\'as a la Tierra en una bola de nieve, \cite{ivan2}.

\section{Proyecto CLOUD}
El experimento CLOUD \cite{ivan5}, realizado en el CERN en Ginebra,  consiste en una c\'amara de acero inoxidable de 3m de di\'ametro que contiene aire ultra puro humidificado y trazas de gases seleccionados,  se coloca en la trayectoria de haces  de piones  que simula los rayos c\'osmicos ionizantes.   Los investigadores que participan en la colaboraci\'on de la CLOUD,  han dado a conocer los primeros resultados de su experimento, ha sido dise\~nado para imitar las condiciones de la atm\'osfera de la Tierra. Su funcionamiento radica  en la disipaci\'on de haces de part\'iculas de sincrotr\'on por medio del acelerador de protones cuya c\'amara se encuentra llena de gas, han descubierto que los rayos c\'osmicos podr\'ian tener un papel determinante en el clima, en particular, en  la producci\'on de aerosoles que potencialmente define la siembra de nubes. 
Hay un gran debate sobre el posible papel de los rayos c\'osmicos en la formaci\'on de estos aerosoles. Henrik Svensmark, del Instituto Nacional del Espacio en Copenhague y sus colegas, postulan la hip\'otesis de que los iones cargados se forman como rayos c\'osmicos que a su vez atraviesan la atm\'osfera y act\'uan como una especie de pegamento, esto  hace que sea m\'as f\'acil para las mol\'eculas unirse y formar aerosoles. Esta hip\'otesis ha sido objeto de controversia, ya que sugiere un papel para la variaci\'on de la radiaci\'on solar, as\'i como las emisiones humanas de gases de efecto invernadero y el cambio clim\'atico, \cite{ivan5}. 
 Para la colaboraci\'on de la investigaci\'on en el proyecto CLOUD, un grupo internacional liderado por Jasper Kirkby donde hace parte del proyecto CERN. El experimento, que ha estado funcionando desde finales de 2009, indica  que cuando se simula la atm\'osfera en s\'olo un kil\'ometro sobre la superficie de la Tierra, el \'acido sulf\'urico, agua y amon\'iaco, crecen en sus componentes  para iniciar la producci\'on de aerosoles. Sin embago, las observaciones  aseguran que dichas componentes no son suficientes  para generar los aerosoles. Estos  valores est\'an por debajo de un factor de hasta mil, incluso cuando los   piones est\'an presentes. Por ello,  llegaron a la conclusi\'on de que otras mol\'eculas tambi\'en deben desempe\~nar un papel, en forma de compuestos org\'anicos. 

Como explica Kirkby, si la sustancia que falta es causada por el hombre, entonces la contaminaci\'on humana podr\'ia estar teniendo un efecto de enfriamiento m\'as grande del que se cree en la actualidad (las emisiones de di\'oxido sulf\'urico son ya conocidas, a trav\'es de las cuales generan el \'acido sulf\'urico, que es vital para la producci\'on de aerosoles). De lo contrario, dice Kirkby, si la sustancia faltante proviene de una fuente natural, el hallazgo podr\'ia implicar la existencia de un nuevo mecanismo de retroalimentaci\'on clim\'atica (posiblemente, las temperaturas m\'as altas ante el aumento de las emisiones org\'anicas de los \'arboles).  Sin embargo, cuando se simula la atm\'osfera superior, los investigadores encontraron un efecto de rayos c\'osmicos m\'as fuerte. Ellos descubrieron que a altitudes de 5 km o m\'as, donde las temperaturas est\'an por debajo de -2$5\,  ^{0}$C, el \'acido sulf\'urico y el agua pueden forman f\'acilmente aerosoles estables, de unos pocos nan\'ometros de di\'ametro y que los rayos c\'osmicos puede aumentar la tasa de producci\'on de aerosol en un factor de 10 o m\'as, \cite{ivan5}.
\begin{figure}[h!]
\centering
\includegraphics[width=120mm]{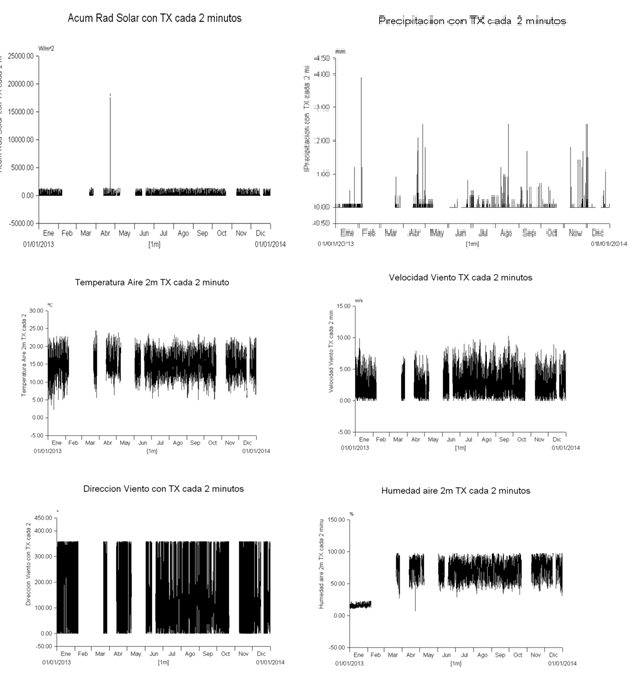}
\caption{Graficas suministradas por el departamento de Meteorolog\'ia y automatizaci\'on, \cite{ivan6}. }
\label{ivan1}
\end{figure}
\section{Mediciones en Colombia}

A lo amplio de la extensi\'on del territorio Colombiano existe entidades de car\'acter privado y oficial que monitorean las variables medioambientales; dentro de ellas est\'a  el Instituto de Hidrolog\'ia, Meteorolog\'ia y Estudios Ambientales (IDEAM), que cuenta con una red de aproximadamente 3000 estaciones (hidrol\'ogicas y meteorol\'ogicas entre convencionales y autom\'aticas), instaladas en diferentes regiones del pa\'is para medir temperatura, humedad, viento, radiaci\'on solar, lluvias, niveles de r\'ios, entre otros, las cuales son operadas a trav\'es de 11 sedes en el pa\'is: Medell\'in, Santa Marta, Villavicencio, Neiva, Barranquilla, Duitama, Pasto, Bucaramanga, Cali, Ibagu\'e y Bogot\'a  sumado a estaciones donde se localizan plataformas de lanzamiento de globos aerost\'aticos. El IDEAM es la entidad nacional encargada de analizar y cuantificar los gases de efecto invernadero (GEI). El IDEAM suministra informaci\'on hidrol\'ogica y meteorol\'ogica que contribuye a garantizar la seguridad a\'erea en el pa\'is.
Hacia 1850 las zonas nevadas que hoy se conocen contaban con un \'area glaciar de 349 $km^2$, \'area que a la fecha se redujo en cerca de 90\% especialmente durante los \'ultimos 20 a\~nos, quedando en cerca de 45.3 $km^2$.
Colombia en cerca de 40 a\~nos ya no tendr\'ia nevados. Con datos generados a finales de 2012, cuenta con cerca de 45.3$km^2$   de masa glaciar distribuida en seis nevados. Cabe destacar que la temperatura aumenta en promedio $0.12^{0}$C por d\'ecada.
El IDEAM adelant\'o un proyecto con excelentes resultados para la recuperaci\'on de masas glaciares en el pa\'is. Instal\'o una valla interceptora de nieve en el Nevado de Sana Isabel, con el fin de acumular nieve en zonas severamente afectadas por el deshielo, \cite{ivan7}.     
La presencia del IDEAM en temas de conservaci\'on estrategias de formaci\'on y sensibilizaci\'on, gesta proyectos que  en cierta medida logran mitigar el impacto social y ambiental ante el cambio clim\'atico, como  poner en marcha junto con el gobierno Colombiano el Proyecto Piloto Nacional de Adaptaci\'on al Cambio Clim\'atico (INAP). 
En las figuras \ref{ivan1}, \ref{ivan2} y \ref{ivan3}, se muestran los  histogramas del a\~no  2013-2014,  variables meteorol\'ogicas recolectadas por el globo aerost\'atico, por parte del (IDEAM). En las figuras \ref{ivan4} y  \ref{ivan5} se muestran los histogramas del primer semestre a\~no de 2014.

\begin{figure}[h!]
\centering
\includegraphics[width=80mm]{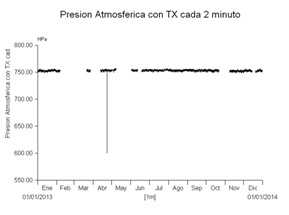}
\caption{Graficas suministradas por el departamento de Meteorolog\'ia y automatizaci\'on, \cite{ivan6}. }
\label{ivan2}
\end{figure}
\begin{figure}[h!]
\centering
\includegraphics[width=120mm]{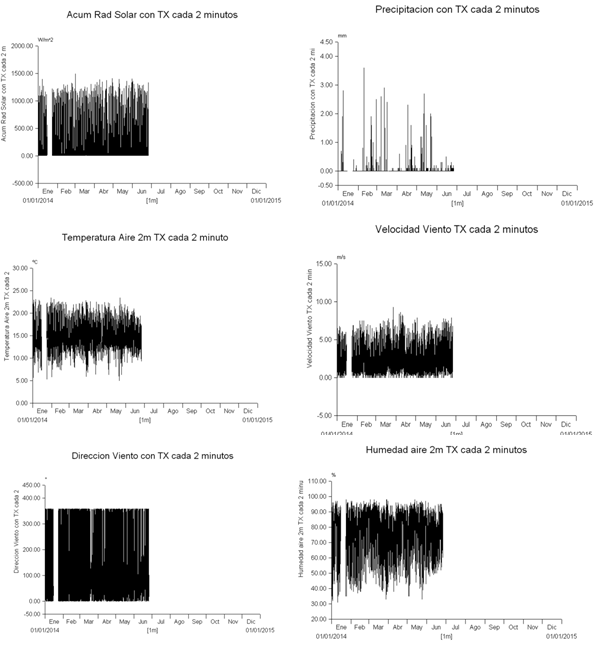}
\caption{Graficas suministradas por el departamento de Meteorolog\'ia y automatizaci\'on, \cite{ivan6}.}
\label{ivan3}
\end{figure}
\begin{figure}[h!]
\centering
\includegraphics[width=80mm]{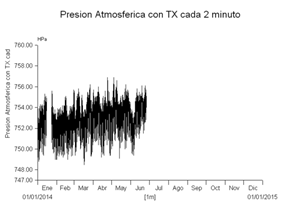}
\caption{Graficas suministradas por el departamento de Meteorolog\'ia y automatizaci\'on, \cite{ivan6}. }
\label{ivan4}
\end{figure}
\begin{figure}[h!]
\centering
\includegraphics[width=80mm]{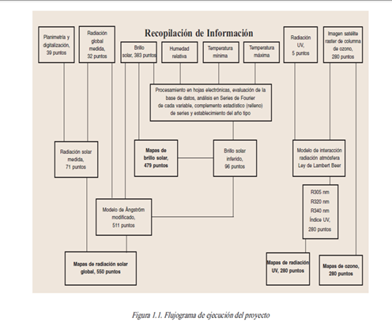}
\caption{Recolecci\'on de informaci\'on para la construcci\'on del atlas solar. Figura tomada de Atlas de radiaci\'on solar de 
Colombia (UPME) e (IDEAM),  2005. }
\label{ivan5}
\end{figure}
Al correlacionar los datos de cada una de las variables sea de radiaci\'on con precipitaci\'on, radiaci\'on con temperatura, temperatura con presi\'on, temperatura con velocidad del viento, obtendremos comportamientos din\'amicos de los fluidos naturales, herramientas que ayudan para originar modelos estad\'isticos que convergen en un dato final de pron\'ostico ambiental.

En las anteriores graficas se observan los datos recolectados en una zona espec\'ifica (aeropuerto el dorado) que hace parte de una estaci\'on, para la construcci\'on del atlas de radiaci\'on por parte de (UPME) e (IDEAM). Cabe resaltar que se requiere informaci\'on de todas las estaciones que se encuentran distribuidas a lo amplio del territorio nacional, labor que en el momento se est\'a realizando; la gesta de este proyecto suministrar\'a informaci\'on relevante para el medio ambiente nacional y en consecuencia global, ya que contribuye al conocimiento de la disponibilidad de los recursos renovables e identifica las regiones estrat\'egicas donde es m\'as adecuada la utilizaci\'on de la energ\'ia solar.  Con esto se podr\'ia dar  soluci\'on de necesidades energ\'eticas; a su vez aporta  repercusiones sobre la vida humana, los ecosistemas y los materiales, al contribuir elementos para prevenir sobre los efectos nocivos de la radiaci\'on terrestre, resultados semejantes los podemos apreciar hoy ante la anterior creaci\'on del atlas solar de 2005 en apoyo de la informaci\'on de textos y modelos del atlas solar de 1993.   La radiaci\'on electromagn\'etica proveniente del sol se propaga radialmente en el espacio vacio; su intensidad disminuye con el cuadrado de la distancia, y su comportamiento se describe empleando las ecuaciones de Maxwell de la teor\'ia electromagn\'etica o mediante la teor\'ia qu\'antica y relativista \cite{ivan6}.
La obtenci\'on del dato de radiaci\'on solar global, se realiza a trav\'es del dataloger que recolecta la informaci\'on  de las mediciones que registra los instrumentos pirheli\'ometro y el piran\'ometro. Estos datos se operan por medio de m\'etodos estad\'isticos como lo son la correlaci\'on o regresi\'on lineal y se interpolan para generar las curvas de calibraci\'on, por normalizaci\'on el instrumento patr\'on nacional es el pirheli\'ometro y es el referente para la calibraci\'on del piran\'ometro. Para la informaci\'on que se logre obtener ante la actual alternativa de aplicaci\'on para la radiaci\'on de fondo  en la radiaci\'on atmosf\'erica, se operar\'ian con una funci\'on de correlaci\'on donde la variable de radiaci\'on global se determina de manera normalizada como se viene trabajando y la variable de radiaci\'on de fondo se operara bajo procesos gaussianos, que define  la convergencia del dato, para posteriormente conseguir una curva te\'orica bajo un modelo de correlaci\'on con los valores de los datos observados; y de este modo obtener el pron\'ostico de la radiaci\'on atmosf\'erica.
   
As\'i como lo fue para el atlas de radiaci\'on solar del 2005 como para el de hoy en d\'ia, se conto con la participaci\'on de la comunidad cient\'ifica resaltando la labor del profesor e Ingeniero Ge\'ografo Simbaqueva Fonseca Ovidio que actualmente se encuentra liderando procesos de investigaci\'on solar en los laboratorios de la Universidad Los Libertadores con la cooperaci\'on  del profesor Ing. Mec\'anico M. Sc. Paguatian Edisson Hernando, \cite{ivan6}.

En la figura \ref{ivan5} se muestra  un diagrama donde se puede observan detalladamente cada uno de los procesos para la recopilaci\'on de informaci\'on desarrollados para el atlas de radiaci\'on de 2005 (UPME) e (IDEAM).
Finalmente,  el Ministerio del Medio Ambiente (2001) tambi\'en ha formulado un programa para el manejo sostenible y restauraci\'on de ecosistemas de la alta monta\~na colombiana en el que se proponen diversas formas de investigaci\'on, conservaci\'on y recuperaci\'on de los pasajes y ecosistemas paramunos a trav\'es de la participaci\'on activa de la colectividad cient\'ifica y de las poblaciones locales y residentes en las \'areas problem\'aticas, as\'i como tambi\'en a trav\'es de la integraci\'on de ecorregiones estrat\'egicas a  trav\'es de trabajos conjuntos entre diversos pa\'ises, ver el art\'iculo \cite{ivan7} para mayor informaci\'on.

 \section{Conclusiones}
En este art\'iculo se present\'o  el Atlas de radiaci\'on de (UPME) e (IDEAM).  En la formulaci\'on estad\'istica se contempla  la mayor cantidad de las variables posibles que inciden en el sistema; para resolver con las magnitudes escalares existentes la variable indeterminada y la distribuci\'on de probabilidad y as\'i la incertidumbre que en nuestro caso es el valor de radiaci\'on terrestre.  Podemos concluir que la informaci\'on (muestreo) requiere del dato de radiaci\'on de fondo con su respectivo intervalo de confianza, con la finalidad de poder correlacionarlo  con las variables que actualmente se trabajan y as\'i originar los modelos estad\'isticos que convergen en un dato final de pron\'ostico ambiental.

Basado en los estudios encaminados a la radiaci\'on de fondo y su potencial importancia en t\'erminos ambientales se logra inferir,  que para la creaci\'on del mapa de radiaci\'on de (UPME) e (IDEAM) se puede emplear una metodolog\'ia que cuente con un  programa de investigaci\'on de radiaci\'on de fondo terrestre, donde se incluya los m\'etodos bayesianos, identificando el modelo correcto para la gesta de este proyecto de investigaci\'on que suministrara informaci\'on relevante para el medio ambiente nacional y en consecuencia global, ya que contribuir\'a  al conocimiento de la disponibilidad de los recursos renovables e identificara las regiones estrat\'egicas donde es m\'as adecuada la utilizaci\'on de la energ\'ia solar, para la soluci\'on de necesidades energ\'eticas; a su vez aportara el nivel de repercusiones sobre la vida humana, los ecosistemas y los materiales, al contribuir elementos para prevenir sobre los efectos nocivos de la radiaci\'on terrestre.

\renewcommand{\refname}{Bibliograf\'ia}
\bibliographystyle{harvard}
\bibliography{Ivan}

\end{document}